\documentclass[twocolumn,showpacs,aps,prl,article]{revtex4}

\usepackage{epsfig}% Include figure files
\usepackage{multirow}
\usepackage{amsmath}

\begin{document}

%\preprint{}

\title{Entropy scaling and thermalization in hadron-hadron collisions at LHC}

\author{Supriya Das}
\email[]{supriya@bosemain.boseinst.ac.in}
\author{Sanjay K. Ghosh}
\email[]{sanjay@bosemain.boseinst.ac.in}
\author{Sibaji Raha}
\email[]{sibaji@bosemain.boseinst.ac.in;sibajiraha@bic.boseinst.ernet.in}
\author{Rajarshi Ray}
\email[]{rajarshi@bosemain.boseinst.ac.in}

\affiliation{Department of Physics, Bose Institute, 93/1 A.P.C. Road, Kolkata - 700009, India \\
and \\ Centre for Astroparticle Physics \& Space Science, Bose Institute, Block EN, Sector V, Kolkata - 700091, India}

%\affiliation{Centre for Astroparticle Physics \& Space Science, Bose Institute, Block EN, Sector V, Kolkata - 700091, India}
\date{\today}

\begin{abstract}
The recent LHC data have once more brought the issue of the formation of 
thermalized state of matter in hadron-hadron collisions into the forefront. 
In this letter, we have shown that the scaling of the information entropy of the chaotically produced 
particles is valid up to recently available data from p+p collisions at 
$\sqrt{s}$ = 2.36 TeV obtained by ALICE experiment. 
We predict that at the highest energies projected for LHC,
almost all the particles will be produced chaotically, indicating that a
collective behavior should be observed in hadronic collisions, as indicated by one of 
present authors quite some time ago.

\end{abstract}

\pacs{12.38Mh, 13.85.-t, 25.75.-q}
%\keywords{}

\maketitle

%\section{}
%\subsection{}
%\subsubsection{}

% If in two-column mode, this environment will change to single-column
% format so that long equations can be displayed. Use
%\begin{widetext}
%\end{widetext}
The properties of strongly interacting matter at extreme conditions have
been an intense area of research for quite some time now. The results from
relativistic heavy-ion collision experiments \cite{raha_physrep,hwa} have 
enriched our knowledge of matter at high temperatures and small densities.
One of the main goals of these experiments is to test our understanding of
Quantum Chromodynamics, the fundamental theory and look for the formation
of deconfined quark-gluon matter.

The non-perturbative nature of QCD makes it difficult to understand all
aspects of particle production in strongly interacting systems. The huge
number of particles produced in heavy-ion collision experiments makes it
even more difficult to decipher the proper sources of these particles.
Under such circumstances, one can look at the simpler systems, such as 
hadron-hadron collision to gain better understanding of particle production 
mechanisms and their possible sources.

Though the research on the multi-particle productions in different systems,
such as high energy hadron-hadron, hadron-nucleus and nucleus-nucleus 
collisions are being pursued by various groups for a long time (see {\it {e.g.}} 
\cite{weiner-IJMPE} and references therein), a proper theory of 
multiparticle production remains elusive till date. More so, as 
non-perturbative nature of the interactions involved makes it difficult to
calculate the soft processes. One of the possible routes to the 
understanding of multiplicity is to investigate the equation of state 
of the matter produced in these collisions \cite{plumer1,plumer2}. This, 
along with a hydrodynamic picture to take care of the overall energy-momemtum
conservation during the space-time evolution of the excited matter, can 
give some idea of the dependence of multiplicity on energy. 

Multiplicity distribution of hadrons produced in high energy particle 
collisions has long been known to deviate from a Poisson distribution 
and has been regarded as a potentially useful source of information about
the underlying production processes. Multiplicity distributions have been
predicted using various models of the low -$p_t$ hadron production 
processes in hadron-hadron collisions \cite{pokorski, groot, capella, 
kaidalov, barshay}. 

The Feynman scaling hypothesis led to the idea of KNO scaling \cite{kno}.
However it may be noted that the inelastic hadronic interactions do not strictly
follow the KNO scaling \cite{wroblewski,thome}.  Inelastic cross section 
is usually made up of single diffraction dissociation and the non-single
diffractive (NSD) part, the latter being the main source of particles produced.
The collision energy dependence of the NSD multiplicity was nicely explained by this 
scaling up to the ISR energies. But this scaling law broke down 
\cite {ua51, ua52} once the collision energy reached the higher SPS energies. 

Analysis of the multiplicity data using statistical moments 
\cite{ua51,ua52} and scaling laws provide new ideas to interpret the 
results. A new quantity, the entropy of multiplicity distributions 
\cite{simak,fowler}, was proposed to revive the scaling law in high energy 
hadronic collisions at the collider energies. An important observation
was the dependence of the multiplicity distribution 
on energy and rapidity bins based on the assumption of the 
existence of two kinds of sources \cite{fowler,giovanini}. Firstly, the chaotic source, which is 
concentrated at small rapidities and has the characteristics of thermally 
equilibrated system and secondly, the coherent source, that contributes to the whole 
rapidity region. It was also shown
that the new scaling, based on entropy, holds good for the entire range 
of energy starting from ISR ($\sqrt{s}$ = 19 GeV) up to the highest 
available collision energy at SPS ($\sqrt{s}$ = 900 GeV) only if the 
entropy is calculated from the chaotically produced particles 
\cite{carruthers}.  
With increasing energy, the weight of the chaotic or 
thermally equilibrated source is expected to increase. This may also be 
indicative of the production of QGP, one of the main goal of all these 
investigations. Recently, deconfinement in hadronic collisions
at TeV energies has been explored using 1-d hydrodynamic model by 
E735 collaboration \cite{porile}. 

In this letter, we will extend the results from this scaling law \cite{carruthers}
for the 
NSD multiplicity distributions in p-p collisions at the highest available 
collision energies at LHC and try to extrapolate to the highest energy projected 
at the LHC. We shall argue that even in hadron-hadron collisions, 
one can expect a thermalised state of matter at sufficiently high 
collision energies, as was already alluded to in \cite{plumer1,plumer2}.

The scaling variable, {\it information entropy} has been calculated within
the context of the two component model \cite{fowler} from multiplicity
distribution. According to this model the emission of particles occurs from
a convolution of a chaotic source with Planck-Polya distribution and a 
coherent source with Poisson distribution. The entropy for symmetric 
rapidity intervals $|\eta|\leq \eta_c$ , for some arbitrary $\eta_c$,
and at center of mass energy $\sqrt{s}$ is given by,
 
\begin{eqnarray}
\nonumber S(\eta_c,\sqrt{s}) 
\nonumber &=& (\langle n_{ch}(\eta_c,\sqrt{s})\rangle +1)~\rm{ln}(\langle n_{ch}(\eta_c,\sqrt{s})\rangle +1) \\
&-&  \langle n_{ch}(\eta_c,\sqrt{s})\rangle~\rm{ln} \langle n_{ch}(\eta_c,\sqrt{s})\rangle
\end{eqnarray}

\noindent
where $n_{ch}$ is the chaotic fraction of the total multiplicity. (Similarly, $n_{co}$
and $n_{tot}$ denote the coherent and the total multiplicity, respectively, so that
$n_{tot} = n_{ch} + n_{co}$.) $n_{ch}$ may be obtained as a product of the total 
multiplicity with the chaotic fraction $\tilde{P}$ as, 

\begin{equation}
\langle n_{ch} \rangle = \tilde{P}\langle n_{tot} \rangle 
\end{equation}

\noindent
where the chaotic fraction is given by,

\begin{equation}
\tilde{P} = [  C_2 - (1+1/\langle n_{tot} \rangle) ]^{1/2}
\end{equation}

\noindent
and $C_2 = \langle {n_{tot}}^2 \rangle/\langle n_{tot} \rangle^2$ is the second moment of 
multiplicity distribution. A scaling law then emerges if we plot the 
scaled entropy $S/\eta_{max}$ as a function of the scaled variable 
$\xi = \eta_c/\eta_{max} $ where 

\begin{equation}
\eta_{max}=\rm{ln}[(\sqrt{s} - 2m_\pi)/m_n] 
\end{equation}

\begin{figure}[htbp]
\includegraphics[scale=0.7]{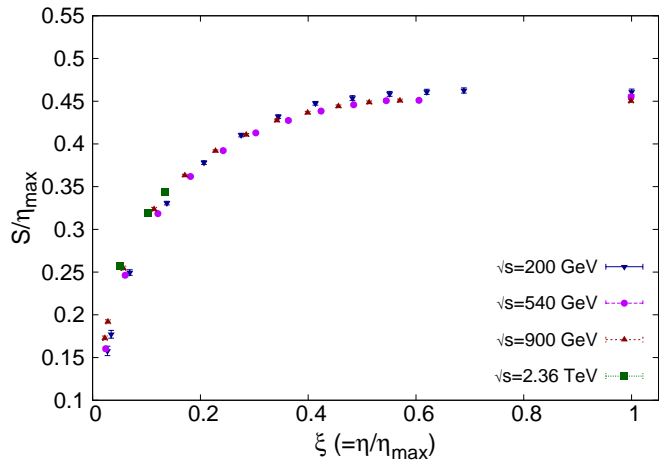}%
\caption{\label{fig:enscaling}Entropy scaling from two component model. 
}
\end{figure}

In Fig.~\ref{fig:enscaling} we have plotted $S/\eta_{max}$ as a function 
of $\xi$ for different collision energies starting from SPS \cite{ua51,ua52} to the 
highest available energy at LHC \cite{alicedata}. 
 As can be seen, all the experimental data follow a single curve within the estimated 
statistical errors. A single
scaling law is thus able to describe multi-particle productions up to the
highest energy available for hadron-hadron collisions.

Expecting that the scaling law holds at even higher energies, an important 
consequence is that we can predict both the average multiplicity as well 
as the nature of the particles produced at higher energies. For this one 
has to extrapolate the available data  to the desired $\sqrt{s}$ for both 
the average total multiplicity $\langle n_{tot} \rangle$ and the average chaotic multiplicity
$ \langle n_{ch} \rangle $. For the extrapolation in $\sqrt{s}$, we choose those three
values of $\xi$ for which experimental data are available for the highest 
$\sqrt{s}$=2.36 TeV at ALICE. Experimental data are not available at
exactly these values of $\xi$ at lower values of $\sqrt{s}$. They are
first obtained through interpolation from the experimental data for $ \langle n_{tot} \rangle $ 
and the model data for $ \langle n_{ch} \rangle $.

\begin{figure}[htbp]
\includegraphics[scale=0.7]{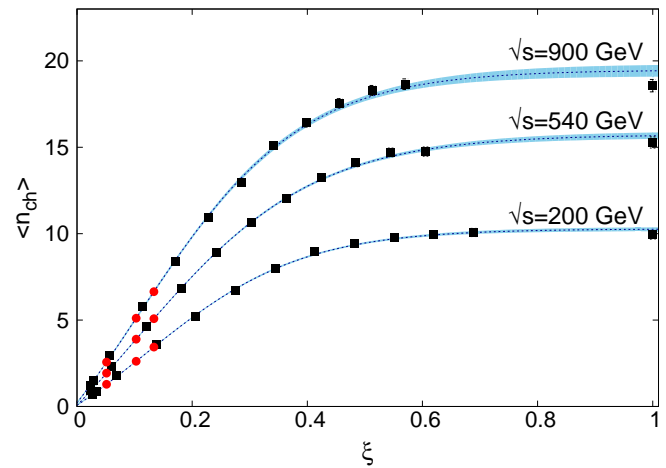}%
\caption{\label{fig:nch_roots}$\langle n_{ch} \rangle$ as a function of 
$\xi$. The squares and circles correspond to data and
interpolated points respectively and the bands show the limits of 
statistical errors. The dotted line is from the fit described in 
the text.}
\end{figure}

In Fig.~\ref{fig:nch_roots} we have plotted the $\langle n_{ch} \rangle$ 
as a function of $\xi$ for different collision energies. A good fit is
obtained for different collision energies with the logistic function,

\begin{eqnarray}
 \langle n_{ch} \rangle = a+\frac{b-a}{1+{\rm e}^{-(\xi-c)/d}}
\label{eqn:nch_xi}
\end{eqnarray}
\noindent
 where $a$, $b$, $c$ and $d$ are parameters to be fitted.
With this fit, we could obtain the value of $\langle n_{ch} \rangle$ at any 
value of $\xi$ for a given $\sqrt{s}$. Thus we could obtain the $ \langle n_{ch} \rangle $
at the desired values of $\xi$. 

\begin{figure}[htbp]
\includegraphics[scale=0.7]{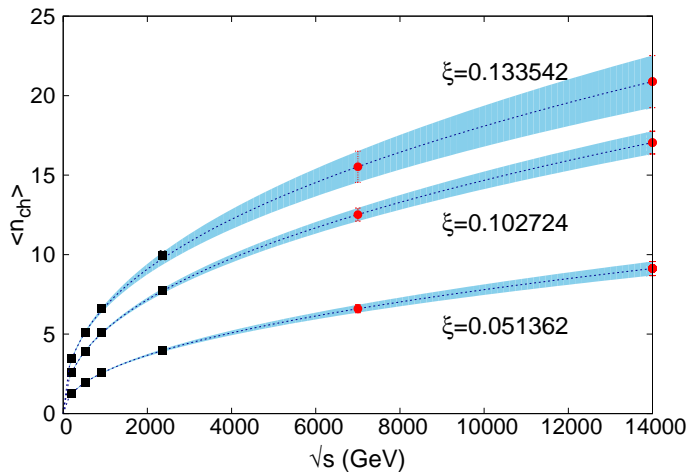}%
\caption{\label{fig:nch}$\langle n_{ch} \rangle$ as a function of collision 
energy. The squares and circles correspond to data and 
predicted points respectively and the bands show the limits of 
statistical errors. The dotted line is from the fit described in 
text.}
\end{figure}

We can now plot the mean multiplicity of chaotically produced particles 
$\langle n_{ch} \rangle$, as a function of collision energy for different 
values of $\xi$ (Fig.~\ref{fig:nch}) obtained from eqn.~\ref{eqn:nch_xi}
and ALICE data at $\sqrt{s}$=2.36 TeV. 
Once the new scaling law for entropy has been established it follows that
$ \langle n_{ch} \rangle $ behaves as a power law with $\sqrt{s}$ for large values
of $ \langle n_{ch} \rangle $. Indeed a good fit is obtained with the formula, 

\begin{eqnarray}
\langle n_{ch} \rangle = b\sqrt{s}~^c
\label{eqn:nch_roots}
\end{eqnarray}

\noindent
where $b$ and $c$ are parameters to be fitted.
The fit parameters are given in the Table~\ref{table:nch_roots}. Using this
fitting function one can now predict the value of $\langle n_{ch} \rangle$ 
at higher collision energies that are expected to be obtained at LHC in 
near future.

%----------------------------------------------------------------------------%
\begin{table}[htbp]
%\caption{Fit parameters of $n_{ch}$ with $\sqrt{s}$ for different $\xi$}
\begin{tabular}{|c|p{.1\textwidth}|p{.13\textwidth}|}
\hline
%----------------------------------------------------------------------------%
$\xi$ & 
\centering Parameters & 
\centering Correlations \tabularnewline
\hline
%----------------------------------------------------------------------------%
\multirow{2}{*}{0.051362} & 
$\begin{aligned}[t]
b &= 0.10(1) \\
c &= 0.47(2) \\
\end{aligned}$ & 
$\begin{aligned}[t]
(b,c) &= -0.991\\
\end{aligned}$\\
\hline
%----------------------------------------------------------------------------%
\multirow{2}{*}{0.102724} & 
$\begin{aligned}[t]
b &= 0.24(2) \\
c &= 0.45(1) \\
\end{aligned}$ & 
$\begin{aligned}[t]
(b,c) &= -0.990\\
\end{aligned}$\\
\hline
%----------------------------------------------------------------------------%
\multirow{2}{*}{0.133542} & 
$\begin{aligned}[t]
b &= 0.35(5) \\
c &= 0.43(2) \\
\end{aligned}$ & 
$\begin{aligned}[t]
(b,c) &= -0.994\\
\end{aligned}$\\
\hline
%----------------------------------------------------------------------------%
\end{tabular}
\caption{Fit parameters of $\langle n_{ch} \rangle$ with $\sqrt{s}$ for different $\xi$}
\label{table:nch_roots}
\end{table}
%----------------------------------------------------------------------------%

\begin{figure}[htbp]
\includegraphics[scale=0.7]{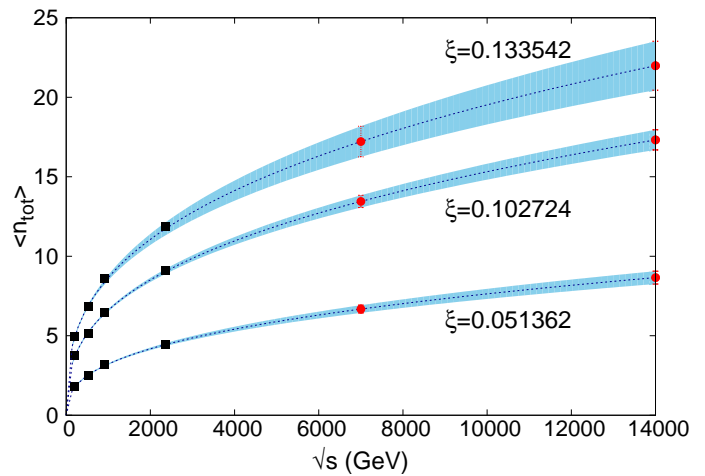}%
\caption{\label{fig:ntot}$\langle n_{tot} \rangle$ as a function of 
collision energy. The squares and circles correspond to
data and predicted points respectively and the bands show the limits
of statistical errors. The dotted line is from the fit described in
text.}
\end{figure}

A similar exercise for the average total multiplicity $ \langle n_{tot} \rangle $ 
(Figure.~\ref{fig:ntot}) gives the collision energy dependence of the average total 
multiplicity for different $\xi$. We again find the best fit to have a
similar functional form as in eqn.~\ref{eqn:nch_roots}. The fit parameters
are shown in Table~\ref{table:ntot_roots}.

%----------------------------------------------------------------------------%
\begin{table}[ht]
%\caption{Fit parameters of $n_{tot}$ with $\sqrt{s}$ for different $\xi$}
\begin{tabular}{|c|p{.1\textwidth}|p{.13\textwidth}|}
\hline
%----------------------------------------------------------------------------%
$\xi$ &  
\centering Parameters & 
\centering Correlations \tabularnewline
\hline
%----------------------------------------------------------------------------%
\multirow{2}{*}{0.051362} & 
$\begin{aligned}[t]
b &= 0.24(3) \\
c &= 0.37(2) \\
\end{aligned}$ & 
$\begin{aligned}[t]
(b,c) &= -0.991\\
\end{aligned}$\\
\hline
%----------------------------------------------------------------------------%
\multirow{2}{*}{0.102724} & 
$\begin{aligned}[t]
b &= 0.53(4) \\
c &= 0.36(1) \\
\end{aligned}$ & 
$\begin{aligned}[t]
(b,c) &= -0.991\\
\end{aligned}$\\
\hline
%----------------------------------------------------------------------------%
\multirow{2}{*}{0.133542} & 
$\begin{aligned}[t]
b &= 0.8(1) \\
c &= 0.35(2) \\
\end{aligned}$ & 
$\begin{aligned}[t]
(b,c) &= -0.995\\
\end{aligned}$\\
\hline
%----------------------------------------------------------------------------%
\end{tabular}
\caption{Fit parameters of $n_{tot}$ with $\sqrt{s}$ for different $\xi$}
\label{table:ntot_roots}
\end{table}
%----------------------------------------------------------------------------%

\begin{figure}[htbp]
\includegraphics[scale=0.7]{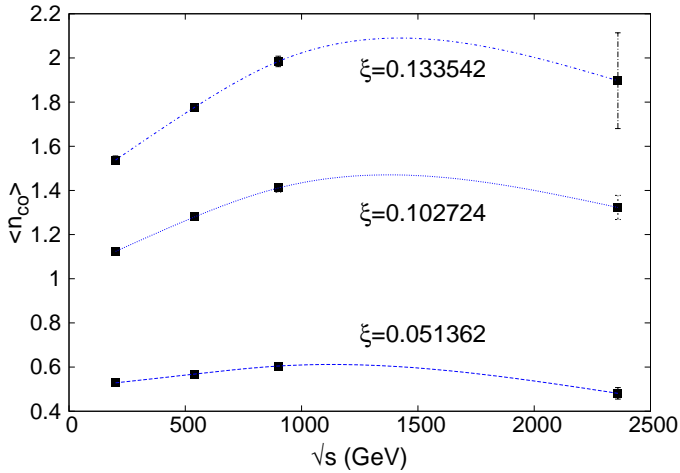}%
\caption{\label{fig:nco}$\langle n_{co} \rangle$ as a function of collision
energy. Dotted line is to guide the eye.}
\end{figure}

The predicted values of both $ \langle n_{ch} \rangle $ and $ \langle n_{tot} \rangle $ 
are marked on the respective figures Fig.~\ref{fig:nch} and Fig.~\ref{fig:ntot}, and
listed in Table~\ref{table:pred}. The table also shows the
contribution of the multiplicity of the coherent source $ \langle n_{co} \rangle $. 
At these high energies the $ \langle n_{co} \rangle $ is consistent with zero. The
nature of $ \langle n_{co} \rangle $ is shown in Fig.~\ref{fig:nco}. We see that the 
$\langle n_{co} \rangle$, after an increase at the collision energies 
up to 900 GeV, indeed shows a dip towards higher energies. The four
points are not enough to obtain a suitable functional form to fit these data. 
The line is just a guide to the eye.

This analysis suggests that at the higher energies the chaotic part of 
multiplicity completely dominates over the coherent fraction. Data at 
higher collision energies that would be available shortly would test these
prdictions.

%----------------------------------------------------------------------------%
\begin{table}[ht]
%\begin{tabular}{|c|c|p{.1\textwidth}|p{.13\textwidth}|}
\begin{tabular}{|c|c|c|c|c|}
\hline
%----------------------------------------------------------------------------%
$\sqrt{s}$ & \multirow{2}{*}{$\xi$} & 
\centering \multirow{2}{*}{$\langle n_{tot} \rangle$} & 
\centering \multirow{2}{*}{$\langle n_{ch} \rangle$} & 
\centering \multirow{2}{*}{$\langle n_{co} \rangle$}  \tabularnewline
%----------------------------------------------------------------------------%
(GeV)& & & & \\
\hline
%----------------------------------------------------------------------------%
\multirow{3}{*}{7000} & 0.051362 & 6.7(2) & 6.6(2) & 0.1(3) \\
 & 0.102724 & 13.4(4) & 12.5(4) & 0.9(6)\\
 & 0.133542 & 17(1) & 16(1) & 1(1)\\
\hline
%----------------------------------------------------------------------------%
\multirow{3}{*}{14000} & 0.051362 & 8.7(4) & 9.1(4) & -0.4(6) \\
 & 0.102724 & 17.3(6) & 17.0(7) & 0.3(9) \\
 & 0.133542 & 22(2) & 21(2) & 1(1) \\
\hline
%----------------------------------------------------------------------------%
\end{tabular}
\caption{Different multiplicities with varying $\xi$ for different $\sqrt{s}$.}
\label{table:pred}
\end{table}
%----------------------------------------------------------------------------%
%----------------------------------------------------------------------------%

To summarise, we have established the scaling law for the information entropy 
obtained in a two source model, as a function of $\sqrt{s}$ upto the maximum
possible range available at present. The multiplicity 
of chaotically produced particles has been evaluated from a two-source model in hadronic 
collisions starting from lowest SPS energy to the highest available 
LHC energy. Our study leads to the prediction that at the highest energies projected for LHC
almost all the particles are produced chaotically {\it i.e.} the 
chaoticity fraction $\tilde{P}$, reaches almost unity, indicating the formation of a thermalized source
and the possibility of observation of collective behaviour in hadronic collisions.


\end{document}